# Multicolor time-resolved upconversion imaging by Adiabatic Sum Frequency Conversion

Michael Mrejen[1,2,†], Yoni Erlich[1,2,†], Assaf Levanon[1,2] & Haim Suchowski[1,2,*]

[1]*Condensed Matter Physics Department, School of Physics and Astronomy, Tel-Aviv University, Tel-Aviv 6779801, Israel*

[2]*Center for Light-Matter Interaction, Tel-Aviv University, Tel-Aviv 6779801, Israel*

[†]*these authors have contributed equally to this work*

*Corresponding author: haimsu@tauex.tau.ac.il*

**Upconversion imaging, where mid infrared (IR) photons are converted to visible and near IR photons via a nonlinear crystal and detected on cheap and high-performance Silicon detectors, is an appealing method to address the limitations of thermal sensors which are expensive, often require cooling and suffer from both limited spectral response and limited spatial resolution as well as poor sensitivity. However, phase matching severely limits the spectral bandwidth of this technique therefore requiring serial acquisitions in order to cover a large spectrum. Here we introduce a novel upconversion imaging scheme covering the mid IR based on adiabatic frequency conversion. We present a mid IR multicolor imaging and demonstrate simultaneous imaging on a CMOS camera of radiation spanning a spectrum from 2 to 4 $\mu m$. Our approach being coherent and ultrafast in essence, we further demonstrate spectrally resolved spatio-temporal imaging which allows spatially distinguishing the temporal evolution of spectral components.**

# 1 Introduction

Spectrally resolved Mid-infrared (Mid-IR) imaging, in which one can visualize information in the fingerprint optical regime [1], is a central tool for remote chemical identification and therefore has applications in chemical [2], biological [3–5], mineralogical [6], and environmental sciences [7] owing to the mid-infrared signatures of vibrionic states in characteristic organic and chemical compounds. Yet, the current technologies for mid-IR imaging [8], such as Indium Antimonide (InSb) and Mercury Cadmium Telluride (MCT), are expensive to process, often require cooling, suffer from noise and poor sensitivity and lack the spectral response and spatial resolution of their VIS-NIR counterpart (Fig.1 (a)). Upconversion imaging [9–12], where mid-IR is converted into visible and near-IR via an optical nonlinear process and detected on cheap and high-performance Silicon sensors, is an appealing method to address these limitations. Many methods have been explored so far, including nonlinear upconversion [9] and photochemical upconversion by triplet states [13]. More recently, nonlinear conversion of incoherent images has been demonstrated, achieving single-photon sensitivity, high resolution imaging of mid-IR signals with a visible camera [14, 15]. However, like any conventional nonlinear optical conversion, these exciting methods suffer from the phase-mismatch challenge, i.e. the lack of inherent momentum conservation between the interacting waves, which requires an implementation of a phase matching compensation technique [16] (Fig.1(b)). This limitation results in an efficient conversion only over a very narrow spectral band, where the phase mismatch can be compensated, and therefore these methods are unsuited for broad-spectral imaging and ultimately require serial acquisitions in order to convert a large spectrum [10,14,15,17].

In recent years, a new concept in nonlinear optics has been developed to tackle the problem of efficient conversion of broad optical bands. The concept, which is known today as adiabatic frequency generation (AFG), was pioneered in the past decade [18, 19], and allows efficient, robust and scalable transfer of broadband, visible and near-IR lasers to the mid-IR optical regime and vice versa. In the past few years this method has been successfully demonstrated and was shown to outperform the currently available mid-IR ultrashort sources, allowing the realization of high-energy, octave-spanning mid-wave IR pulsed source that can reach up to sub two-cycle temporal resolution in the mid-IR covering the 2-5 $\mu$m spectral region [20–22]. Yet, until today this technique has been applied only to the generation of coherent sources and has not been used as an imaging platform.

Here, we introduce a novel upconversion broadband imaging scheme based on adiabatic frequency conversion [23, 24] to convert ultra-broadband mid-IR coherent images into the visible-near-IR spectral range and demonstrate broadband color imaging spanning one octave in the mid-IR (2-4 $\mu$m) using low-cost, high sensitivity, fast, and robust visible CMOS, all without having to tune the converting crystal phase matching conditions (Fig.1 (c)). Furthermore, based on our ultrafast scheme, we also show spectrally resolved spatio-temporal imaging using ultrafast sub-picosecond gating allowing the observation, on the picosecond scale, of the spectral evolution of the image in the mid-IR at two simultaneous wavelengths (2 and 4 $\mu$m). Besides the temporal aspect, this constitutes a proof concept for simultaneous imaging of a broadband mid-IR scene. We also discuss the parameters that affect the spatial resolution performances of the proposed scheme. Our method ultimately offers a way towards full spectrally resolved single-shot imaging of noncoherent mid-infrared scenes with Silicon based sensors with applications in remote sensing for chemical, biological and industrial applications as well as fundamental scientific implications with the direct spatio-temporal characterization of exotic materials such as 2D and layered materials.

## 2  Experimental methods

We first designed an Adiabatic Sum Frequency Generation (ASFG) crystal, where the phase mismatch between the pump (1030 nm) and the upconverted mid-IR signals (2-4 $\mu m$) is compensated by using the poling method, resulting in an adiabatically varying period (from 14.4µm to 23.8µm) along the crystal's propagation axis which provides an average of 20% conversion efficiency over the 2-4 $\mu m$ range (more details on the design and conversion efficiency can be found in the Supplementary material).

Next, we incorporate our ASFG crystal in the imaging setup, outlined in Fig. 2(a), enabling ultrabroad upconversion imaging from the mid-IR (2 $\mu$m to 4 $\mu$m) into the VIS-NIR spanning from 690 nm to 820 nm. The apparatus is composed of a 2 MHz repetition rate ultrafast pulse at 1030 nm with 20 nm full width half-maximum (FWHM), which is used as a pump to generate tunable mid-IR radiation in the 2-4 $\mu m$ range via a tunable OPG crystal (Fig.2(a)) as well as to pump the upconversion ASFG process. The generated mid-IR illuminates an USAF 1951 resolution target mask imaged in a modified 4-$f$ system with the ASFG crystal placed at the Fourier plane, upconverting the mid-IR into the VIS-NIR via the 1030 nm pump. The VIS-NIR image is then formed on a color Silicon focal plane array after proper filtering of the 1030 nm pump and parasitic second harmonic generation (515 nm). A delay line allows the tuning of the temporal overlap between the pump and the mid-IR (a more detailed setup description is available in the Supplementary material).

## 3 Results and Discussions

We experimentally demonstrate upconversion imaging over the full 2 $\mu$m to 4 $\mu$m span detected on a color CMOS camera. In Fig. 2(b), the mid-IR wavelength images: 2, 3.3 and 4$\mu m$ are upconverted to 680, 790, 820 nm images, respectively. In contrary to standard upconversion imaging approaches, our method does not require changing anything in the experiment setup due to the broad acceptance of the adiabatic crystal. We note that by using the Bayer filter of the CMOS camera we are able to resolve the different VIS-NIR colors. Owing to the conversion process, these colors have a one-to-one relationship to the mid-IR wavelength and therefore these results constitute the first demonstration of multicolor mid-IR imaging based on ASFG upconversion without phase-matching compensation. We emphasize that a shorter pump wavelength (typically 720 nm for which powerful CW laser diodes are widely available) will allow shifting the upconverted signal to the blue, taking full advantage of ASFG designed spectral coverage (1500 nm 5000 nm) and of the RGB decomposition of the Silicon sensor.

We further show the ultrafast temporal imaging capabilities of our method. As seen in Fig.3 (c, inset), we start from an OPG mid-IR signal comprised of two distinct wavelengths at 2 and 4 $\mu m$. The generated mid-IR is further passed through a 5-mm thick Silicon window that, due to its dispersion, causes temporal separation between the mid-IR spectral components as can be observed in the retrieved mid-IR temporal profile in Fig.3(a) (orange curve). This mid-IR radiation illuminates a standard calibration target and is imaged in the same ASFG based imaging setup described above. This time, however, the delay line is scanned and VIS-NIR images (Fig.3 (b)) and spectra (depicted in a 2D map Fig.3 (c)) are recorded simultaneously as a function of the delay between the 1030 nm pump and the mid-IR signal. As can be seen, the Silicon window induces enough spectral dispersion in the mid-IR two-color signal to be temporally distinguishable by the relatively short 1030 nm pump (∼ 800 fs), allowing the imaging of the same mask at the two wavelengths separately (Fig.3 (c), top and bottom dashed lines). Owing to the unique efficient broadband conversion capability of the ASFG, simultaneous upconversion of the two wavelengths is also observed (as shown in Fig.3 (c) middle dashed line), as a result of overlap of the 1030 nm pump and the two temporally dispersed wavelengths. A control experiment with a stretched 1030 nm pump (∼ 2 ps) and no Silicon window is presented in the Supplementary material, showing no ability to temporally separate the two spectral components.

It is worth noting the different magnification for the two wavelengths, especially noticeable in the middle line case. We attribute this effect to two main reasons. First of all, there is an inherent scaling of the image in a manner that relates to the generated visible light and the upconverted mid-IR wavelength [10], that goes as a ratio between the wavelengths and focal lengths at play $I_{up} \sim I_{obj}(-\frac{\lambda_1 f_5}{\lambda_2 f_6}x, -\frac{\lambda_1 f_5}{\lambda_2 f_6}y)$, where $I_{obj}$ is the intensity distribution at the object plane, $\lambda_1$ is the mid-IR wavelength, $\lambda_2$ is the upconverted wavelength and $f_5, f_6$ the focal lengths involved in the 4-$f$ system as described in the Supplementary Material. From this relation we observe that for the different wavelengths in the mid-IR, and their associated VIS-NIR wavelengths, the scaling will be different. We also note the lateral chromatic aberration that can be accounted for by the wide span of different angles coming from the object as shown in Fig.3(b) in the lines region at the edge of the field of view. Finally, inherent to the ASFG process 23, the VIS-NIR wavelengths are generated at different locations along the optical axis at the ASFG crystal, resulting in different locations of the beams focus. A careful optimization of the optics and achromatism will result in the same magnification for the full mid-IR bandwidth and all the colors will overlap.

We further investigate our scheme performance by testing the resolving power of our imaging system on a standard 1951 USAF target. As we show in the Fig.4, the ultimate resolution we achieve is at 2 $\mu m$ with $28.51 \frac{lines\ pairs}{mm}$, in agreement with the theoretical achievable resolution as detailed in the supplementary material. It is important to note that the main factor currently limiting the resolution is the small size of crystal aperture (1 mm in width, only 1 mm height- current standard fabrication capability) which acts as a spatial filter in the Fourier plane and results in filtering out the high spatial frequencies since the Fourier Transform distribution is larger than the crystal aperture [14]. While this is a fundamental limitation, it can be easily overcome as we perfect the nonlinear crystal fabrication process, that can, in principle, achieve much larger sizes in height and width [25].

Finally, we would like to note that while we have demonstrated here imaging on RGB channels on a CMOS camera, there are methods to retrieve continuous hyperspectral data, after calibration, from RGB sampled data[26]. Our ASFG method, by preserving faithfully the spectral shape of the converted spectrum[20,22], provides the grounds for such retrieval and therefore opens the door for single shot hyperspectral mid-IR imaging. Such data will provide a FTIR-like spatially resolved spectral data and will allow in further investigations molecular identification.

In conclusion, we have shown in this work that by harnessing the adiabatic conversion scheme for upconversion imaging, we are able, in a single shot process, to upconvert mid-IR images spanning from 2 *μm* to 4 *μm* into the visible-near-IR. This achievement, enabled by the inherent broadband and efficient frequency conversion of the adiabatic approach, outperforms current phase-matching limited upconversion imaging methods known to date. Furthermore, we demonstrate spectrally resolved spatio-temporal imaging capabilities as well as simultaneous, multi-color imaging of mid-IR scenes onto a color CMOS camera. This method paves the way for high-resolution, and ultrafast capabilities in mid-IR remote sensing and spatio-temporal characterization of exotic classes of materials such as 2D and layered materials.

## 4 Materials and Methods

Detailed Materials and Methods can be found in the Supplementary Material.

## 5 Acknowledgements

This work has been funded by the European Research Council under the grant MIRAGE 20/15.

# 6 Competing Interests

The authors declare that they have no competing financial interests.

# 7 Contributions

H.S. and M.M. conceived the original idea. Y.E. designed the adiabatic crystal. M.M. designed the experimental setup. M.M. and Y.E. carried out the experiments and analysed the data. A.L. established the pump-probe apparatus. H.S. provided guidance and supervision. All authors participated to the manuscript preparation.

ciples, concepts and applications in plant tissue analysis. *Molecules* **22** (2017). URL https://www.mdpi.com/1420-3049/22/1/168.

5. Shi, J. *et al.* High-resolution, high-contrast mid-infrared imaging of fresh biological samples with ultraviolet-localized photoacoustic microscopy. *Nature Photonics* **285**, 1 (2019).

6. Pieters, C. M. & Englert, P. A. J. *Remote Geochemical Analysis, Elemental and Mineralogical Composition* (Cambridge University Press, 1993).

7. Blunck, D. L., Gross, K. C. & Rhoby, M. R. Mid-IR hyperspectral imaging of laminar flames for 2-D scalar values. *Optics Express* **22**, 21600–21617 (2014).

8. Vollmer, M. & Möllmann, K.-P. *Infrared Thermal Imaging*. Fundamentals, Research and Applications (Wiley-VCH Verlag GmbH & Co. KGaA, Weinheim, Germany, 2017).

9. Boyd, R. W. & Scully, M. O. Efficient infrared imaging upconversion via quantum coherence. *Applied Physics Letters* **77**, 3559–3561 (2000).

10. Pedersen, C., Karamehmedović, E., Dam, J. S. & Tidemand-Lichtenberg, P. Enhanced 2d-image upconversion using solid-state lasers. *Optics Express* **17**, 20885–20890 (2009).

11. Grisard, A. *et al.* Near-infrared to visible upconversion imaging using a broadband pump laser. *Optics Express* **26**, 13252–13263 (2018).

12. Zhao, L. *et al.* Mid-infrared upconversion imaging pumped by sub-nanosecond micro-cavity laser. In Zhao, Y. (ed.) *Fifth International Symposium on Laser Interaction with Matter*, 1104607 (International Society for Optics and Photonics, 2019).10

**Figure 1:** State-of-the-art mid-IR imaging methods for multispectral object, schematically rendered by the authors' institution (Tel Aviv University logo) where different parts of the logo have different spectral composition. (a) Thermal imaging is the most widely used method for mid-IR sensing. Mercury Cadmium Telluride (MCT) and Indium Antimonide (InSb) based solutions require cooling, and are severely limited in sensitivity and spatial resolution. Furthermore, the resulting images are integrated over a spectral bandwidth and therefore lack color differentiation. (b) Upconversion imaging addresses some of the thermal imaging limitations by converting the mid-IR radiation to visible-near-IR to be subsequently imaged on Silicon based imager. This method is appealing since these detectors are cheap, fast, efficient, high resolution and color sensitive. This technique however is impeded by the phase mismatch, an inherent aspect of nonlinear frequency conversion processes. The standard nonlinear crystal technology results in efficient upconversion only for a narrow spectral band- width. Therefore, a tuning mechanism via angle of incidence (as shown) or temperature needs to be applied to the nonlinear crystal and sequential images. (c) In contrast, Adiabatic Frequency Conversion based upconversion imaging allows efficient ultra-broadband nonlinear conversion capabilities onto a standard Silicon focal plane array.

**Figure 2:** (a) The experimental setup includes an Optical Parametric Generation (OPG) crystal that produces tunable mid-IR radiation spanning from 2$\mu m$ to 4$\mu m$ when pumped with 1030 nm pulse. The generated mid-IR illuminates a mask, which is then imaged through a 4$f$ a system comprised of lenses $f_3$ and $f_4$. Upconversion of the mid-IR image into the Visible-Near IR (VIS-NIR) occurs at the Fourier plane (between $f_3$ and $f_4$) where an Adiabatic Sum

Frequency Generation (ASFG) crystal mixes the mid-IR with the 1030 nm pump to produce the VIS-NIR Fourier transform of the mid-IR mask image. The VIS-NIR upconverted image is then formed with $f_4$ on a Color CMOS Camera image allowing the simultaneous detection of the full span of the upconversion from 680 nm to 820 nm. We emphasize that the unique efficient ultrabroadband conversion allowed by the ASFG process allows this simultaneous imaging without requiring any parameter change in the upconversion process to compensate for phase mismatch. (b) The upconverted images for three different mid-IR spectra that were obtained on the color CMOS camera are presented. We note the different colors rendered by the Bayer RGB filter. (c) The respective original mid-IR spectra obtained from the OPG process. (d) The respective visible spectrum after upconversion showing coverage from $2\mu m$ to $4\mu m$. More technical details on the setup are provide in the Supplementary methods.

**Figure 3:** Time and spectrally resolved mid-IR imaging. (a) Based on the setup detailed in Fig. 2(a), we temporally stretch, using a 5-mm thick Silicon window (noted as "Si" in Fig. 2(a)), a two-color (2μm and 4μm) mid- IR radiation obtained via the OPG process (inset). The stretched mid-IR is used in a similar manner as in Fig.2 and images along with VIS-NIR spectra are recorded as a function the delay between the mid-IR and the 1030 nm pump. The time profile of the mid-IR (orange solid curve) and the 1030 nm pulse (dot black curve) relative position in time are depicted for three data points among the many collected. (b) The resulting VIS-NIR images shows that the mid-IR spectral components have experienced significant dispersion and that 1030 nm, 800 fs pulse can resolve them separately with a time delay of about 1 ps. We observe however that for a time delay between 0.2 to 0.8 ps, where there is a temporal overlap for both spectral components, we are able to obtain an upconverted image that includes both of the mid-IR colors simultaneously. the specifics of the imaging process are further detailed in the text in Fig. 4. (c) The collected VIS-NIR spectra are plotted as function of the time delay. We observe that the Silicon window temporally disperses the two colors and that the pump is narrow enough to resolve the two colors separately.

**Figure 4:** Resolution of the Adiabatic camera at 2 (left column), 3 (middle column) and 4 (right column) $\mu m$. We examine the resolution achieved with the USAF 1951 resolution target for group 4, elements 2- 17.96 $\frac{lines\ pairs}{mm}$ (top row), 3- 20.16 $\frac{lines\ pairs}{mm}$ (middle row) and 6- 28.51 $\frac{lines\ pairs}{mm}$. For the 2 $\mu m$ all the images are clearly observed, in accordance with the calculated resolution derived in the supplementary material. In the 4 $\mu m$ the resolution seems low in all the images but the first one, as predicted, since the maximum calculated spatial frequency is 18.75 $\frac{lines\ pairs}{mm}$. While these figures are far from the mid-IR diffraction limit, the resolution is currently mainly limited by the acceptance aperture of the ASFG crystal that acts as a low pass filter as explained in the main text.

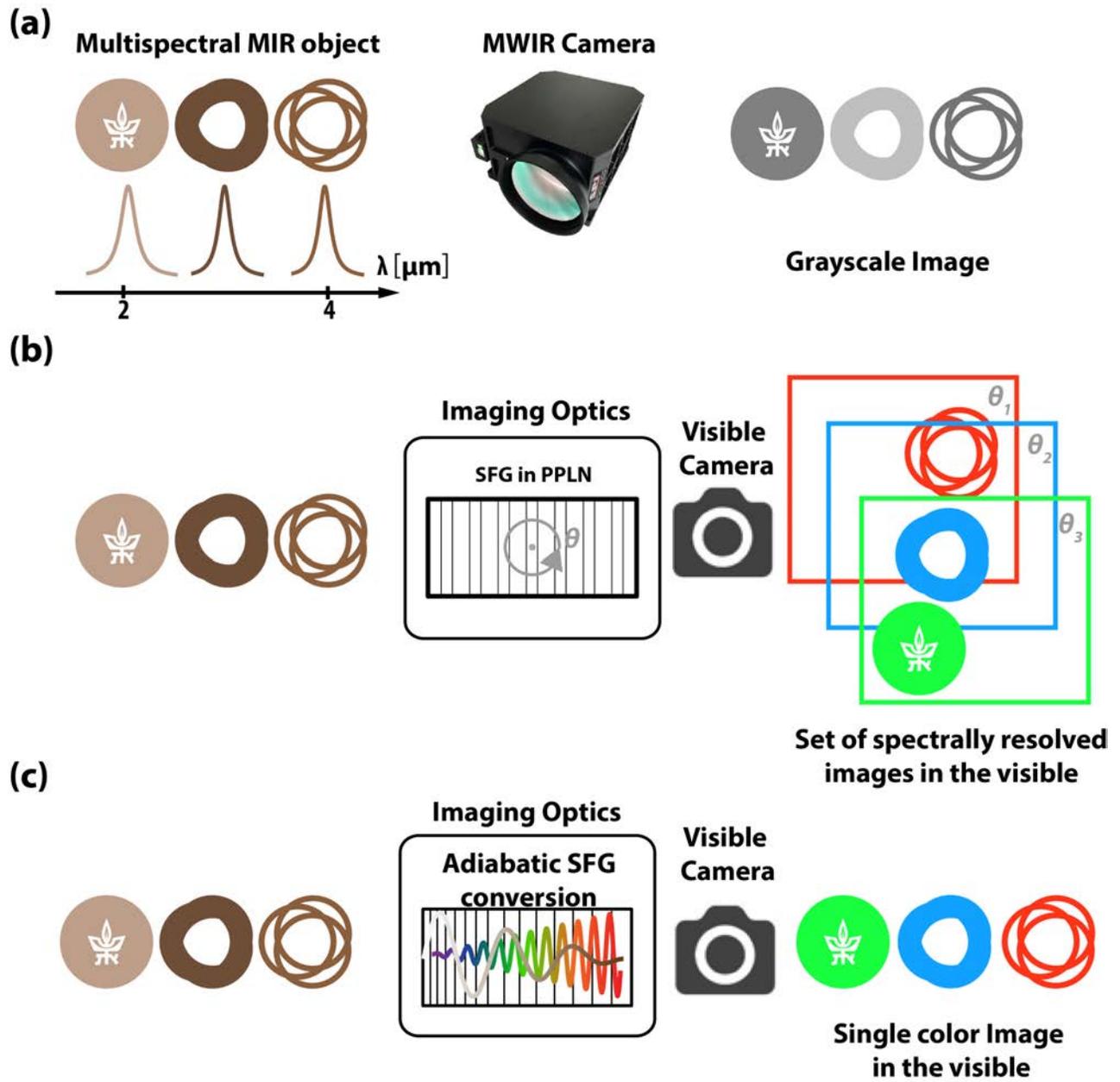

Figure 1

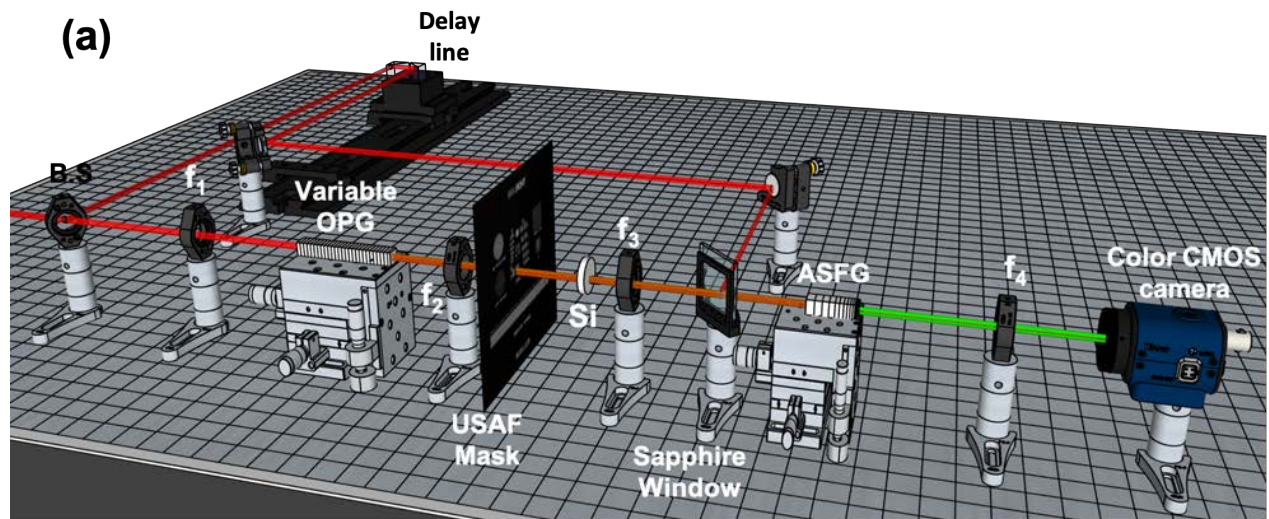

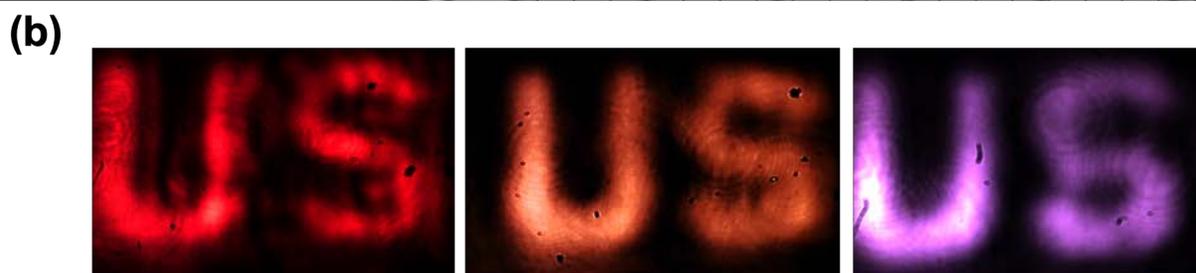

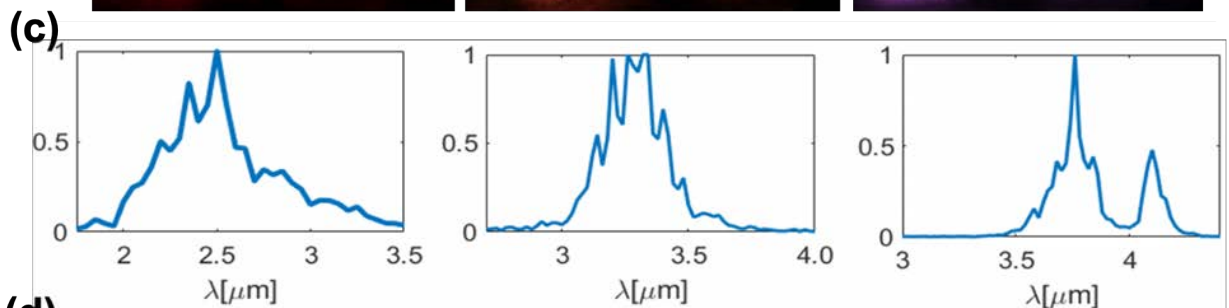

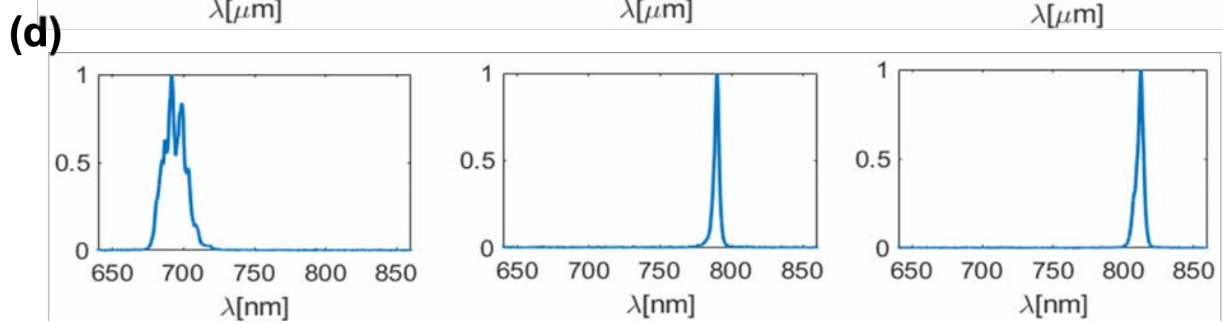

Figure 2

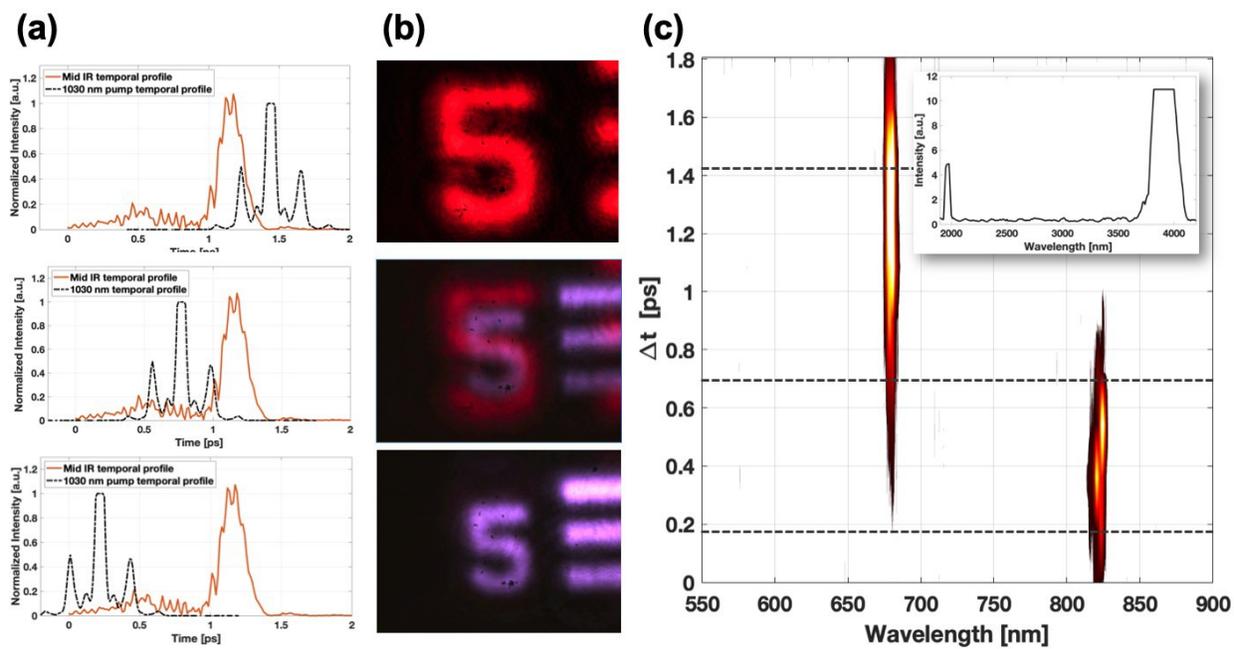

Figure 3

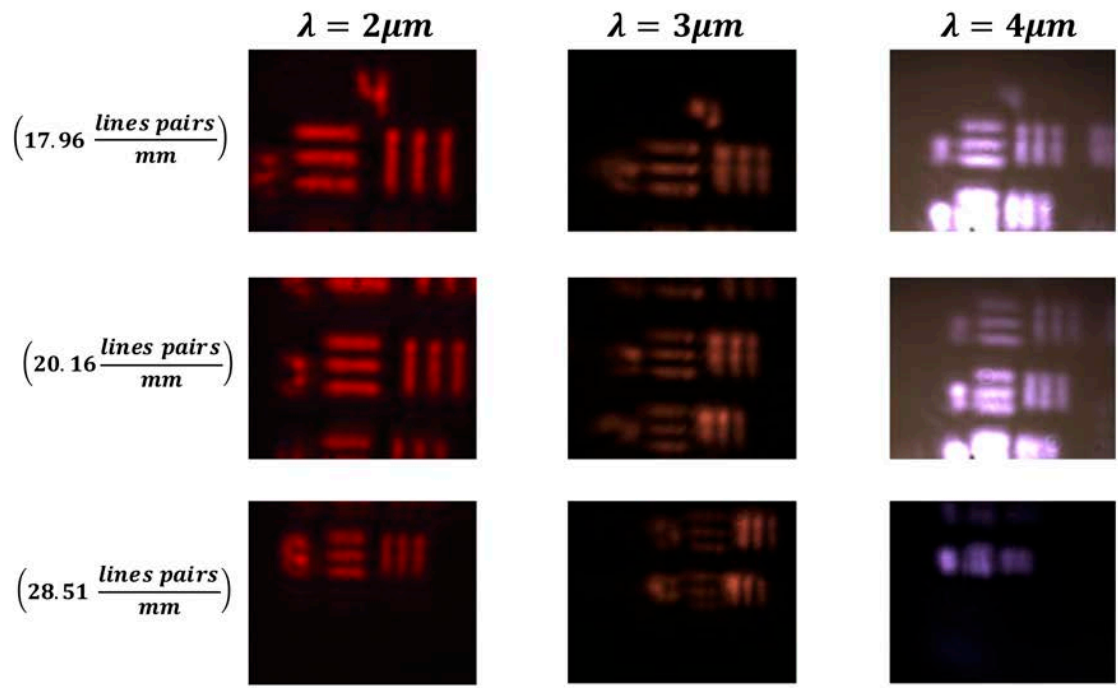

Figure 4